\documentclass[a4paper,12pt]{article}  
\usepackage{amssymb}
\usepackage{graphicx}
\usepackage{amsmath}  
\def \bea {\begin{eqnarray}}  
\def \eea {\end{eqnarray}}  
\def \mea {\nonumber\\}  
 
\def \bea {\begin{eqnarray}}
\def \eea {\end{eqnarray}}
\def \mea {\nonumber\\}
\def \vs {\vskip5mm}
\def \ni {\noindent}
\def \half {{\textstyle \frac{1}{2}}}

\begin{document}

\begin{titlepage}

\title{Massless Dirac particle in a stochastic magnetic field: A solvable quantum walk approximation}     
     
\author{
A.J. Bracken$^{a}$\footnote{{\em Email:} a.bracken@uq.edu.au}, D. Ellinas$^{b}$\footnote{{\em Email:} ellinas@science.tuc.gr} and I. Smyrnakis$^{c}$\footnote{ {\em Email:} smyrnaki@tem.uoc.gr}
\\$^{a}$School of Mathematics and Physics\\    
The University of Queensland\\Brisbane 4072, Australia
\\
\\$^{b}$Department of Sciences,
Division of Mathematics
\\Technical University of Crete
\\Chania, GR 731 00, Greece
\\
\\$^{c}$Technical Education Institute of Crete
\\PO Box 1939 Heraklion
\\  GR 71004, Greece 
}

\date{}     
\maketitle     
\vs\ni
\vs\ni
\newpage     
\begin{abstract}
A massless Dirac particle is considered, moving along the $x$-axis while Pauli-coupled by its anomalous 
magnetic moment to a
piecewise constant magnetic field along the same axis,  with stochastically varying sign.  
The motion is approximated as a quantum  walk with unitary noise, for which the evolution can be found
exactly.  
Initially ballistic, the motion approaches a classical diffusion on a time-scale determined by 
the speed of light, the size of the magnetic moment, the strength of the field
and the time interval between changes in its direction.    
It is suggested that a process of this type could occur in the Sun's corona, significantly affecting the solar fluxes of one or more neutrino types.  
\end{abstract}

     
\end{titlepage}    

\setcounter{page}{3}

\section{Introduction}
It is known that the time-evolution of some free-particle relativistic wave equations can be approximated by a quantum walk (QW).  This connection has its origins in Feynman's path-integral approach to the propagator for Dirac's equation \cite{feynman,meyer}, and so predates the extensive researches into QWs and their potential for applications in quantum information theory that have  
followed from the seminal paper of Aharonov {\em et al.} \cite{aharonov}. 
 
The connection has been explored in work by Childs and Goldstone on the massless Dirac equation 
\cite{childs}, by Katori {\em et al.} \cite{konno} on Weyl's neutrino equation, and by 
Strauch \cite{strauch} and  us  \cite{BES} on Dirac's equation with a non-zero mass.  As a result 
it is now well-understood how the dynamics of a free relativistic 
particle with spin $1/2$\,, moving on a line, can be approximated arbitrarily closely by a 
simple, one-dimensional QW.   

Because the free Weyl and Dirac equations can be solved easily and exactly, the connections with QWs are 
mathematically interesting 
but of limited importance to relativistic physics.  In contrast,  we consider here
a relativistic quantum system that does not seem to be amenable to analysis {\em unless} approximated as a noisy QW.
  
Hackett \cite{hackett} considered 
the effect of adding an arbitrarily small amount of classical randomness 
in a unitary way to a simple QW on the line.  The idea of a QW contaminated by unitary noise
was then explored extensively by Shapira {\em et al.} \cite{biham}.  An initially ballistic 
motion with a spreading rate proportional to the elapsed time, as typical of a QW, 
is eventually replaced by a diffusive motion, with a 
spreading rate proportional to the square-root of the time, as typical of a classical random walk (CRW).  
The transition occurs 
after a time (number of iterations) determined by the strength of the noise.
Addition of noise in other forms to QWs has been shown to produce similar effects \cite{brun,kendon}. 

Recently we have studied a particularly simple one-dimensional QW with unitary noise, with the property that
an initially diagonal density matrix remains diagonal during the evolution of the system \cite{BE2,BE3}. 
The form of this evolving
density matrix has been 
found exactly.   This simple system may be regarded as a toy model for QWs with unitary noise,
and it does exhibit the transition from ballistic to diffusional behavior explicitly.  

What we now show is that, despite its simplicity, 
this model has an application to
the description of a neutral, massless Dirac particle with anomalous magnetic moment, Pauli-coupled to
a piece-wise constant magnetic field with stochastically changing direction, and confined to move on a straight line.  

The interpretation of the transitional behavior of the QW in the context of this relativistic 
system is rather remarkable: at short times $t$ 
there is a high probability of finding the particle at a distance  
$ct$ from its starting point, 
where $c$ is the speed of light, as expected for a free massless particle.  But
as time passes, the motion tends more and more towards a diffusion on the line, 
with a high probability 
of finding the particle within a distance $\alpha\sqrt{t}$ of a certain point $x_0$ as a normal 
distribution is approached,  centered on that point.  Here $\alpha$ is a constant whose value,
together with that of $x_0$, is determined by the speed of light, the 
magnetic moment, the strength of the magnetic field, and the time interval between changes 
in field direction.

\section{The relativistic system}
The dynamics of the particle in this case is governed by Dirac's equation in the form
\bea
i\hbar \,\partial\,\psi({\vec x},t)/\partial\, t=H(t)\psi({\vec x},t)\,,\quad
\mea\mea
{\rm where}\quad H(t)=c{\vec \alpha}\cdot {\vec p}- \mu\,\beta\, {\vec S}\cdot {\vec B}(t)\,.
\label{dirac1}
\eea 
Here  ${\vec p}=-i\hbar \partial/\partial{\vec x}$ is the momentum $3$-vector, $\mu$ is the 
magnetic moment, ${\vec B}(t)$ is the external magnetic field $3$-vector, and $ {\vec S}$ is
the spin $3$-vector.  The $4\times 4$  matrices in \eqref{dirac1} are conveniently defined in 
terms of the $2\times 2$ 
Pauli matrices ${\vec \sigma}$ and $2\times 2$ unit matrix $I_2$ by
\bea
{\vec \alpha}=\sigma_3\otimes {\vec \sigma}\,,\quad \beta=\sigma_2\otimes I_2\,,\quad {\vec S}=\half\,\hbar \, I_2\otimes {\vec \sigma}\,.
\label{matrices2}
\eea

Can such a system be realized in Nature or the laboratory?  We postpone a discussion of this until 
the end of the paper; at this stage our objective is to add to the collection of 
relativistic systems amenable to mathematical  analysis.

To proceed, suppose that $p_{\,2}\,|\psi\rangle=p_{\,3}\,|\psi\rangle= 0$ on states $|\psi(t)\rangle=\psi(x,t)$ of interest,  writing
$x_1=x$, $p_1=p=-i\hbar \partial/\partial x$.   We consider   magnetic fields of the form
${\vec B}(t)=(B(t),\,0,\,0)$ with
\bea
B(t)=sB_0 \,,\quad N\Delta<t<(N+1)\Delta\,,\quad N=0,\,1,\,\dots\,,
\label{Bform}
\eea
where $\Delta>0$ and $B_0$ are constants,
and $s=\pm 1$, each sign having probability $1/2$, independently at each value of $N$.
Under these assumptions, the helicity operator $\half I_2\otimes \sigma_1$ is a constant of motion, 
and we may suppose for definiteness that 
\bea
(I_2\otimes \sigma_1)|\psi\rangle=|\psi\rangle\,,
\label{constant1}
\eea  
and work henceforth in this eigenspace, neglecting the action of the Pauli matrices in the
second factor of the tensor product.
Now the
Hamiltonian   reduces effectively to 
\bea
H=cp\,\sigma_3 - (\hbar\mu B(t)/2)\,\sigma_2\,,
\label{effectiveH}
\eea 
and the evolution operator carrying $|\psi(t=n\Delta)\rangle$ into  $|\psi(t=(n+1)\Delta)\rangle$ reduces to
\bea
V(s)&=& \exp\{-i[cp\sigma_3- s(\hbar\mu B_0/2)\,\sigma_2]\Delta/\hbar\}\,.  
\label{evolution1}
\eea

\section{Approximation as a noisy quantum walk}
The key step in our approximate treatment is to write from \eqref{evolution1}
\bea
V(s)
&\approx & \exp\{-i[cp\Delta/\hbar]\sigma_3\}\,\,\exp\{i[s\mu B_0\Delta /2]\sigma_2\}\,.
\label{evolution2}
\eea
Bearing in mind the Campbell-Baker-Hausdorff formula \cite{miller}, we assume that \eqref{evolution2} 
is a good approximation provided 
\bea
|c\langle p\rangle_n \Delta/\hbar| \ll 1 \quad{\rm and}\quad |\mu B_0\Delta/2|\ll 1\,,
\label{condition1}
\eea
where $\langle p\rangle_n$ is the expectation value of  $p$ in the state $|\psi(t=n\Delta)\rangle$. 
We shall not attempt here a more rigorous analysis of the approximation \eqref{evolution2}, analogous to that given for the 
free-particle \cite{BES}, but content ourselves with the assumption that inequalities \eqref{condition1} hold so strongly 
for each $n=0\,,1\,,2\,,\dots\,,N$ that there is a negligible accumulation of errors when the approximate evolution operators in the form 
\eqref{evolution2} are applied $N$ times to $|\psi (0) \rangle$, while the system evolves
over a time $t=N\Delta$ of interest (see below). 

The operators \eqref{effectiveH}, \eqref{evolution1} and \eqref{evolution2} act on the Hilbert space ${\cal H}={\cal H}_W\otimes {\cal H}_C$, 
where ${\cal H}_W$ is the space of square-integrable 
functions of $x$ on which $p$ acts, and ${\cal H}_C$ is the 2-dimensional complex vector space on which the Pauli matrices in \eqref{effectiveH} and \eqref{evolution1} act,  
each space having the usual scalar product.  
In ${\cal H}_W$ we introduce the infinite sequence of orthonormal 
states $|k\rangle$ defined by 
\bea
|k\rangle=F(x-kc\Delta)\,,\quad k=0\,,\pm 1\,,\pm 2\,,\dots\,,
\label{k_states}
\eea
where $F(x)$ is some chosen smooth function with compact support ${\cal S}<(-c\Delta/2,c\Delta/2)$, satisfying 
\bea
\int_{\cal S} |F(x)|^2\,dx =1\,.
\label{normalize}
\eea
Note that the states $|k\rangle$
do not in general form a basis in ${\cal H}_W$.  We also introduce
on ${\cal H}_W$, the translation operators
\bea
E_{\pm}=\exp{(\mp ic p \Delta/\hbar)}\,,\quad E_{\pm}|k\rangle=|k\pm 1\rangle\,.
\label{shifts}
\eea
In ${\cal H}_C$ we introduce the orthonormal basis of states $|\tau \rangle$, $\tau=\pm 1$, with
\bea
\sigma_3|\pm 1\rangle=\pm|\pm 1\rangle\,,
\label{spin_basis}
\eea
and the associated orthogonal projection operators
\bea
P_{\pm}=\half(I_2\pm\sigma_3)\,,\quad P_{\pm}|\pm 1 \rangle=|\pm 1\rangle\,,\quad  P_{\pm}|\mp 1 \rangle=0\,.
\label{projections}
\eea 

Now we can  rewrite  \eqref{evolution2}
as 
\bea
V(s)=E_+\otimes P_+U(s) + E_-\otimes P_-U(s)\,,
\label{Udef1}
\eea  
where 
\bea
U(s)&=&\exp \{ i[s\mu B_0\Delta /2]\sigma_2 \}
\mea\mea
&=&\cos(\mu B_0\Delta/2)I_2+is\sin( \mu B_0\Delta/2)\sigma_2\,.
\label{Udef2}
\eea
In
\eqref{Udef1} we have introduced $\otimes$ to separate explicitly the spatial and spin degrees of freedom. 
Adopting the standard matrix form for $\sigma_2$, we have from \eqref{Udef2} the matrix representation
\bea
U(s)= {\cal N}
\left[  
\begin{array}{l}  
1\quad\quad\quad s\epsilon\\  
-s\epsilon\quad\quad 1  
\end{array}  
\right]\,,
\label{Udef3}
\eea
where  
\bea
\epsilon=\tan(\mu B_0\Delta /2)\,,\quad {\cal N}=1/\sqrt{1+\epsilon^2}\,. 
\label{epsilon_def}
\eea
Here we have used $\cos ( \mu B_0\Delta/2)\geq 0$, which follows from the 
second of conditions \eqref{condition1}.

At this point we recognize \eqref{Udef1} as the evolution operator for a QW on 
a line with unitary noise \cite{hackett,biham,BE2,BE3}.  
In that context, ${\cal H}_W$ is
the `walker' space and
${\cal H}_C$ the `coin' space, $E_{\pm}$ governs steps by the walker to 
right or left on the line, and $U(s)$ is the `reshuffling matrix'.
The walk is an essentially trivial one, with $U=I_2$,
contaminated by unitary noise that is characterized by the parameter 
$\epsilon$ and the random variable $s$.

\section{Analysis}
Depending on the direction of the magnetic field, the initial  state vector $|\psi(0)\rangle$ evolves to either
\bea
|\psi(\Delta)\rangle=V(+1)\,|\psi(0)\rangle \quad {\rm or}\quad |\psi(\Delta)\rangle=V(-1)\,|\psi(0)\rangle\,,
\label{newstate1}
\eea
at time $t=\Delta$, each possibility occurring with probability $1/2$.  But this simply means that the initial density operator 
$\rho(0)=|\psi(0)\rangle\langle\psi(0)|=\rho_0$, say, 
evolves into
\bea
\rho(\Delta)=\rho_1=\half\sum_{s=\pm 1}V(s)\rho_0 V(s)^{\dagger}\,.
\label{newstate2}
\eea
More generally, at time $t=N\Delta$, the density operator is
\bea
\rho(N\Delta)=\rho_N\qquad\qquad\qquad\qquad
\qquad\qquad\qquad\qquad\qquad\qquad\qquad\qquad
\mea\mea
=\frac{1}{2^N}\sum_{s_1,s_2,\,\dots\,,s_N}
V(s_N)\dots V(s_2)V(s_1)\,\rho_0\,V(s_1)^{\dagger}V(s_2)^{\dagger}\dots V(s_N)^{\dagger}\,,
\label{newstate3}
\eea
each $s_n$ being summed over the values $\pm 1$. 

Consider the case with 
\bea
|\psi(0)\rangle=|k=0\rangle\otimes|\tau=+1\rangle\,,
\label{start_vec1}
\eea
so that 
\bea
\rho_0=|0\rangle\langle 0|\otimes
\left[  
\begin{array}{l}  
1\quad\quad 0\\  
0 \quad\quad 0 
\end{array}  
\right]\,,  
\label{initial4}
\eea
using again the matrix representation as in \eqref{Udef3}. We 
postpone to the end of the paper 
a discussion of the difficulty of finding a positive energy state in this form for the system with Hamiltonian \eqref{effectiveH}. 
In this case,
\bea
V(+1)|\psi(0)\rangle &=&{\cal N}\left(
E_+\otimes
\left[  
\begin{array}{l}  
1\quad\quad \epsilon\\  
0 \quad\quad 0 
\end{array}  
\right]
+E_-\otimes
\left[  
\begin{array}{l}  
0\quad\quad 0\\  
-\epsilon \quad\quad 1 
\end{array}  
\right]
\right)
\mea\mea
&\qquad&\qquad\qquad\times
\left(
|0\rangle \otimes
\left[  
\begin{array}{l}  
1\\  
0 
\end{array}  
\right]\right)
\mea\mea
&=&
{\cal N}\left(|1\rangle\otimes
\left[  
\begin{array}{l}  
1\\  
0  
\end{array}  
\right]
+ |-1\rangle\otimes 
\left[  
\begin{array}{l}  
0\\  
-\epsilon 
\end{array}  
\right]\right)\,.
\label{example1}
\eea
It follows that
\bea
V(+1)\,\rho_0\,V(+1)^{\dagger}={\cal N}^2\left(
|1\rangle\langle 1|\otimes
\left[  
\begin{array}{l}  
1\quad\quad 0\\  
0 \quad\quad 0 
\end{array}  
\right]\right.
\mea\mea
\left.
+
|1\rangle\langle -1|\otimes
\left[  
\begin{array}{l}  
0\quad\quad -\epsilon\\  
0 \quad\quad 0 
\end{array}  
\right]
+
|-1\rangle\langle 1|\otimes
\left[  
\begin{array}{l}  
0\quad\quad 0\\  
-\epsilon \quad\quad 0 
\end{array}  
\right]
\right.
\mea\mea
\left.
+
|-1\rangle\langle -1|\otimes
\left[  
\begin{array}{l}  
0\quad\quad 0\\  
0 \quad\quad \epsilon^2 
\end{array}  
\right]
\right)\,.
\label{example2}
\eea
The expression for $V(-1)\,\rho_0\,V(-1)^{\dagger}$ is similar, with $\epsilon$ replaced by $-\epsilon$ throughout, and it then follows from
\eqref{newstate2} that in this case 
\bea
\rho_1=
{\cal N}(\epsilon)^2\left(
|1\rangle\langle 1|\otimes
\left[  
\begin{array}{l}  
1\quad\quad 0\\  
0 \quad\quad 0 
\end{array}  
\right]\right.
\mea\mea
\left.
+
|-1\rangle\langle -1|\otimes
\left[  
\begin{array}{l}  
0\quad\quad 0\\  
0 \quad\quad \epsilon^2 
\end{array}  
\right]
\right)\,.
\label{example3}
\eea

We see that $\rho_1$, like $\rho_0$, is diagonal in the space of states spanned by all the vectors $|k\rangle\otimes|\tau\rangle$.  
It is easily shown by induction  
that the same is true of $\rho_N$ for each non-negative integer $N$, and that in fact $\rho_N$ has the form
\bea
\rho_N=\sum _{k=-N}^{N} \,\!^{'} |k\rangle\langle k|\otimes 
\left[  
\begin{array}{l}  
\alpha_{N\,k}\quad\quad 0\\  
0 \quad\quad \beta_{N\,k} 
\end{array}  
\right]\,,
\label{example4}
\eea
where the prime indicates that the sum is over the $N+1$ values  $-N\,,-N+2\,,\dots\,,N$ of $k$, and the constants 
$\alpha_{N\,k}$, $\beta_{N\,k}$ are non-negative.  

It is also easily shown \cite{BE2,BE3} from \eqref{newstate3} that $\alpha_{N\,k}$, $\beta_{N\,k}$ are
determined by
the coupled recurrence relations
\bea
\alpha_{N+1\,k}={\cal N}^2\,(\alpha_{N\,k-1}+\epsilon^2\beta_{N\,k-1})\,,\quad\!\!\! k=-N+1\,,-N+3\,,\dots\,,N+1\,,
\mea\mea
\beta_{N+1\,k}={\cal N}^2\,(\beta_{N\,k+1}+\epsilon^2\alpha_{N\,k+1})\,,\quad \!\!\! k=-N-1\,,-N+1\,,\dots\,,N-1\,,
\mea\mea
{\rm for}\quad N=0\,,1\,,2\,,\dots\,,\qquad\qquad\qquad
\label{example5}
\eea
with the  initial conditions
\bea  
\alpha_{0\,0}=1\,,\qquad \beta_{0\,0}=0\,,\quad \alpha_{0k}=\beta_{0 k}=0\,,\quad k\neq 0\,.   
\label{ICs2}  
\eea
For example, \eqref{example3} shows that 
\bea
\alpha_{1\,1}={\cal N}^2\,,\quad \alpha_{1\,-1}=0\,,\quad
\beta_{1\,1}=0\,,\quad \beta_{1\,-1}={\cal N}^2\,\epsilon^2\,.
\label{example6}
\eea

Note that 
\bea
\langle +1|\otimes\langle k|\,\rho_N\,|k\rangle\otimes|+1\rangle =\alpha_{Nk}\,,
\quad \langle -1|\otimes\langle k|\,\rho_N\,|k\rangle\otimes|-1\rangle =\beta_{Nk}\,,
\label{probs1}
\eea
so that $\alpha_{Nk}$ (resp. $\beta_{Nk}$)  is the probability of finding the particle in the state $|k\rangle\otimes |+1\rangle$
(resp. $|k\rangle\otimes |-1\rangle$) on measurement.  Then
\bea
P_{Nk}=\alpha_{Nk}+\beta_{Nk}
\label{probs2}
\eea
is the probability that the particle will be found in the state $|k\rangle$, with the value of $\tau$ immaterial. Because
the state $|k\rangle$ can be localized as closely as we like about $x=kc\Delta$, we may say that $P_{Nk}$ 
is the probability of finding the particle `at' that place, at time $t=N\Delta$.  

When $\epsilon=0$, the solution of \eqref{example5} and \eqref{ICs2} is $\alpha_{NN}=1$, with all other $\alpha_{Nk}$ and all $\beta_{Nk}$
vanishing.  The massless particle is free, and marches to the right at speed $c$, with 
probability $P_{NN}=1$ of being at $x=Nc\Delta=ct$ at time
$t=N\Delta$.  If we had chosen $\tau=-1$ in the initial state, the particle would have marched to the left.  
In each case the motion is always purely ballistic; there is no transition to diffusional evolution at large times, because  
there is no noise
contaminating the QW, which has the trivial reshuffling matrix $U=I_2$.   
If we were to choose an initial mixed state instead of \eqref{initial4}, taking
$\rho_0=|0\rangle\langle 0|\otimes I_2/2$, the resulting probability distribution on the $x$-axis would have $P_{NN}=P_{N\,-N}=1/2$, 
giving an extreme example of the familiar two-horned distributions associated with simple QWs on the line \cite{ambainis}.   

When $\epsilon=1$, \eqref{example5} and \eqref{probs2} show that
\bea
P_{N+1\,k}= \half(P_{N\,k-1}+P_{N\,k+1})\,,
\label{pascal1}
\eea
which is the defining rule for the Pascal's triangle of successive probability 
distributions centered on $k=0$ for a simple classical random walk (CRW),
leading to
\bea
P_{Nk}=\frac{1}{2^N}\,C^N_{(N+k)/2}\,,\quad k=-N\,,-N+2\,,\dots\,,N
\label{pascal2}
\eea
for $N=0\,,1\,,2\,,\dots$, where $C^p_q=p!/q!(p-q)!$ is the binomial coefficient.    
Then \eqref{example5} and \eqref{ICs2} give 
\bea
\alpha_{Nk}=\half P_{N-1\,k-1}=\frac{1}{2^{N}}\,C^{N-1}_{(N+k-2)/2}\,,\quad k=-N+2\,,-N+4\,,\dots N\,,
\mea\mea
\beta_{Nk}=\half P_{N-1\,k+1}=\frac{1}{2^{N}}\,C^{N-1}_{(N+k+2)/2}\,,\quad k=-N\,,-N+2\,,\dots N-2\,,
\mea\mea
\alpha_{N\,-N}=0\,,\quad \beta_{N\,N}=0\,,\quad \alpha_{0\,0}=1\,,\quad \beta_{0\,0}=0\,.\qquad\qquad
\label{pascal3}
\eea
Thus there is no ballistic regime in this case; the evolution is purely diffusional.  
As is well known \cite{hughes}, 
the distribution \eqref{pascal2} is asymptotic to a continuous,  normal one centered 
on $k=0$ as $N\to\infty$, with density
\bea
P(N,k)= \frac{e^{-k^2 /2N}}{\sqrt{2N\pi}}\,,\qquad -\infty<k<\infty\,.
\label{CRW1}
\eea
Then we can say that after a time $t=N\Delta$ with $N\gg 1$, 
there is  $95\%$ probability that the particle can be found within two standard deviations of 
the origin, that is to say, with $|x|<2\sqrt{N} c\Delta=2c\sqrt{t\Delta}$, in sharp contrast 
to the behavior of the free, massless particle, which would always be found at a distance $ct$ 
from the origin.  
  
In terms of the physical variables of the relativistic particle, we have asymptotically as $t=N\Delta\to \infty$, the density
\bea
{\cal P}(t,x)= \frac{e^{-x^2/4Dt}}{\sqrt{4\pi Dt}}\,,\qquad -\infty<x<\infty\,,\quad D= c^2\Delta/2\,.
\label{prob_distbn2}
\eea
Note however that \eqref{condition1} requires $\epsilon\ll 1$, so that our approximate treatment of 
the relativistic system 
may break down when $\epsilon=1$.    
  
For general values of $\epsilon$, the solution of \eqref{example5} and \eqref{ICs2} 
has been found in the form \cite{BE3}
\bea
\alpha_{NN}=Q(N,N)\,,\quad \beta_{NN}=0\,,\qquad\qquad\qquad\qquad
\mea
{\rm and}
\qquad\qquad\qquad\qquad\qquad\qquad\qquad\qquad\qquad\qquad\qquad\qquad\qquad\qquad
\mea
\alpha_{N\,k}= Q(N,k) - Q(N-1,k+1)/(1+\epsilon^2)\,,\qquad\qquad
\mea\mea
\beta_{N\,k}=\epsilon^2 Q(N-1,k+1)/(1+\epsilon^2)\,,\qquad\qquad\qquad
\label{solutions1}
\eea
for $k=-N\,,-N+2\,,\dots\,,N-2$, where
\bea
Q(N,k)=\frac{1}{(1+\epsilon^2)^N}\,\sum_{s=0}^{(N-|k|)/2}\,C^{N-2s}_{(N-k-2s)/2} 
\,C^{N-s}_s \,(\epsilon^4-1)^s\,.
\label{Qdef}
\eea 
Then \eqref{probs2} gives
\bea
P_{NN}=Q(N,N)\,,\quad {\rm and }\quad P_{N\,k}=Q(N,k)-\frac{1-\epsilon^2}{1+\epsilon^2}\, Q(N-1,k+1)
\mea\mea
{\rm for}\quad k=-N\,,-N+2\,,\dots\,,N-2\,.\qquad\qquad\qquad
\label{solutions2}
\eea

As $N=t/\Delta$ increases for any  given $0<\epsilon <1$, the probability distribution \eqref{solutions2} 
undergoes the transition from ballistic evolution to diffusive evolution mentioned 
in the Introduction.  This is  seen in Fig. 1, which shows probability versus $x/c\Delta=k$ in 
the case $\epsilon =0.2$, for $N=t/\Delta=20\,,50\,,100\,,200\,.$ The plots reveal also that during the 
transition, 
the probability distribution may be considered to consist of two components: (1) 
the probability at the point $x=ct$ corresponding to ballistic motion of the massless particle at 
speed $c$, marked by the circled point on the right in each plot, which decreases as time $t=N\Delta$ 
increases, and (2) a growing diffusional distribution on  $x$-values closer to the origin, eventually
swamping the ballistic component, and approaching a normal distribution 
centered on a mean positive displacement.  
\begin{figure}[t]
\centering
\includegraphics[width=10cm]{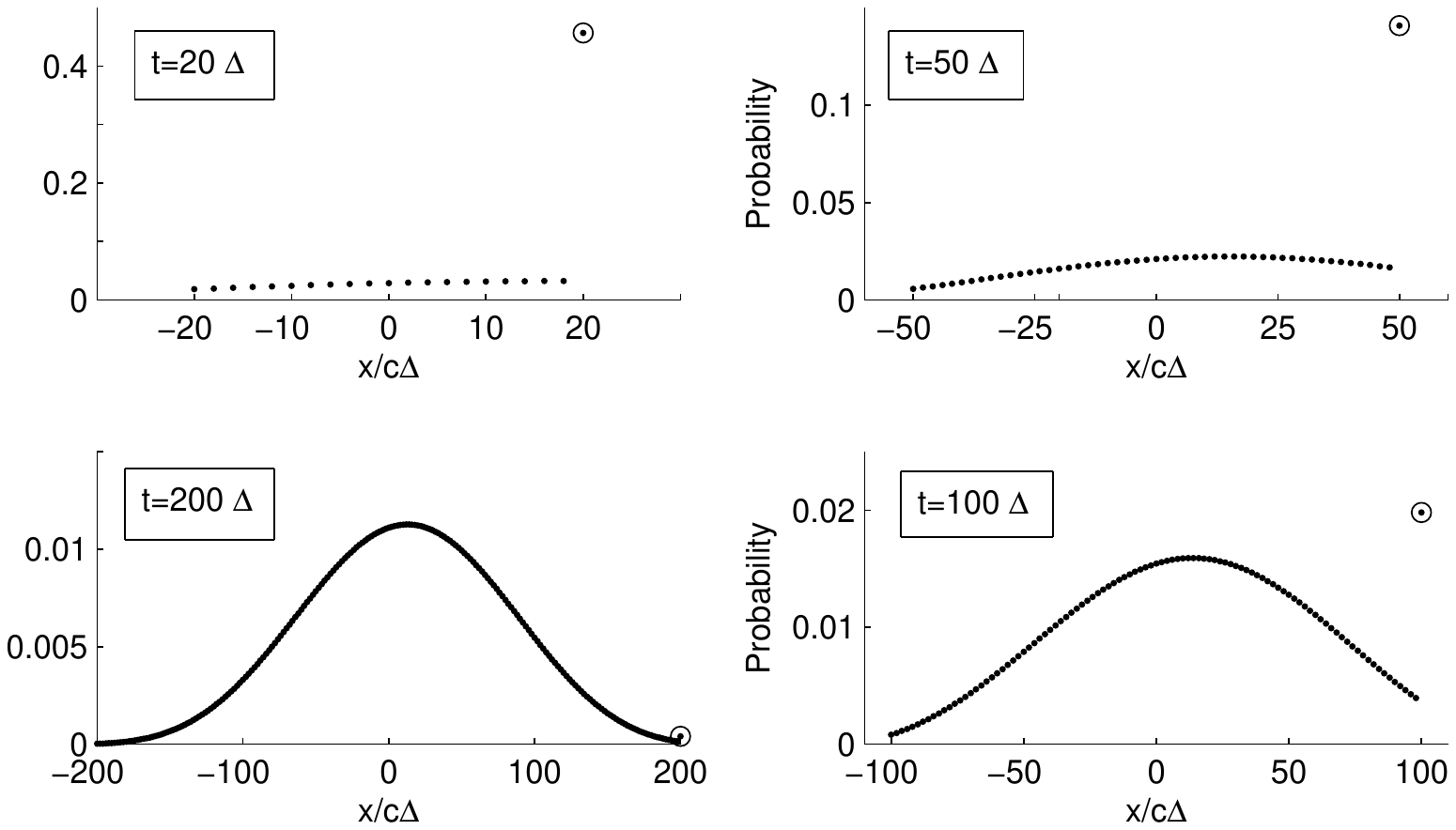}
\caption{Clockwise from Top Left: Plots of probability against $x/c\Delta$, for $t=20\Delta\,,50\Delta\,,
100\Delta \,,200\Delta$, with $\epsilon=0.2$.  The probability at $x=ct$ is marked with a circled dot in each plot.  Note the different scales on the axes in the plots.}
\end{figure}

This transition is reflected in the behavior of $P_{NN}$, the probability that 
the particle is at $x=ct$ at time  $t=N\Delta$ (the circled dot on the plots in Fig. 1).    
This probability equals $1$ in the free particle case $\epsilon=0$. From \eqref{solutions2} and 
\eqref{Qdef} we have in general
\bea
P_{NN}= [1/(1+\epsilon^2)]^N\,C^{N}_{0} \,C^{N}_0 = 1/(1+\epsilon^2)^{N}\,.
\label{extreme_prob1}
\eea
As $N=t/\Delta$ increases, there is  a steady decline in the probability that the particle 
continues to travel at speed $c$ to the right.  This is clear in the successive plots of Fig. 1, and is shown
explicitly  in Fig. 2.
\begin{figure}[t]
\centering
\includegraphics[width=10cm]{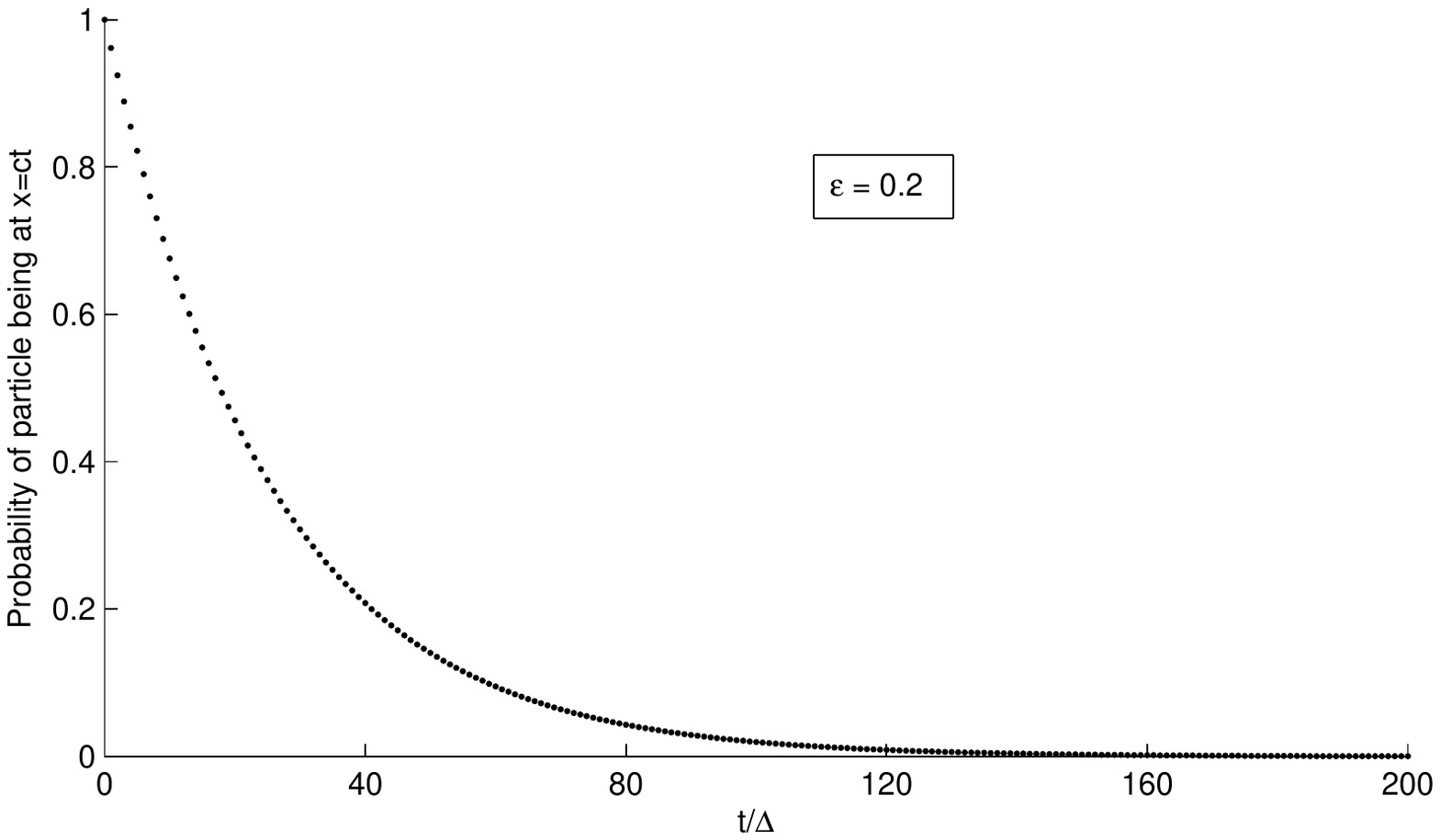}
\caption{Probability that the particle is at $x=ct$ after time $t$, in the case $\epsilon=0.2$.}
\end{figure}

It is also instructive to consider the moments of the probability distribution \eqref{solutions2}, defined 
for each $N$ as  
\bea
S_N^{(p)}=\sum_{k=-N}^N\,\!'\, P_{Nk}\,k^p\,,\quad p=0\,,1\,,2\,,\dots\,.
\label{moments1}
\eea 
The zeroth, first and second moments  have been calculated exactly \cite{BE3},  as 
\bea
S_N^{(0)}=1\,,\quad S_N^{(1)}=\frac{(1-\epsilon^2)}{2\epsilon^2}\,\left[ 1-\left(\frac{1-\epsilon^2}{1+\epsilon^2}\right)^N\right]\,,
\mea\mea\mea
S_N^{(2)}=\frac{1}{2\epsilon^4}\left[2N\epsilon^2-1+\epsilon^4+\frac{(1-\epsilon^2)^{N+1}}{(1+\epsilon^2)^{N-1}}\right]\,.
\label{moments2}
\eea

In the ballistic regime, $N\epsilon^2\ll 1$, and     
\bea  
(1-\epsilon^2)^{N+1}/(1+\epsilon^2)^{N-1}= 1-\epsilon^4-2N\epsilon^2+2N^2\epsilon^4+
{\rm O}([N\epsilon^2]^3)\,,  
\label{expansion1}  
\eea  
so that    
\bea 
S_N^{(2)}=\frac{1}{\epsilon^4}(N\epsilon^2)^2+{\rm O}([N\epsilon^2]^3)\,.
\label{regime2}  
\eea  
The rate of growth of the second moment,   
quadratic in $N$, is characteristic of a QW \cite{ambainis}.   
  
In the diffusive regime, $N\epsilon^2\gg 1$, and    
\bea
S_N^{(2)}= \frac{1}{\epsilon^4}\,(N\epsilon^2)-\frac{1-\epsilon^4}{2\epsilon^4}+{\rm O}(1/[N\epsilon^2])\,,
\label{regime1}  
\eea  
showing the rate of growth is now  linear in $N$, as typical 
of a CRW \cite{hughes}. 
Note that however small is $\epsilon^2$, eventually  
$N$ becomes so large that the second regime is reached.    
The changeover occurs in the region 
where $N\epsilon^2={\rm O}(1)$.  This is  the behavior observed in numerical simulations  
of quantum walks with unitary noise \cite{biham}, 
of which the present process may be considered  
a simple, special case, with the advantage that it is more amenable to mathematical analysis.  

Note also that the second moment can be interpreted as the expectation value
$\langle (x/c\Delta)^2\rangle $.  Fig. 3 shows
the transition from quadratic to linear behavior of this quantity with increasing $N=t/\Delta$ in the case $\epsilon=0.2$. 
\begin{figure}[t]
\centering
\includegraphics[width=10cm]{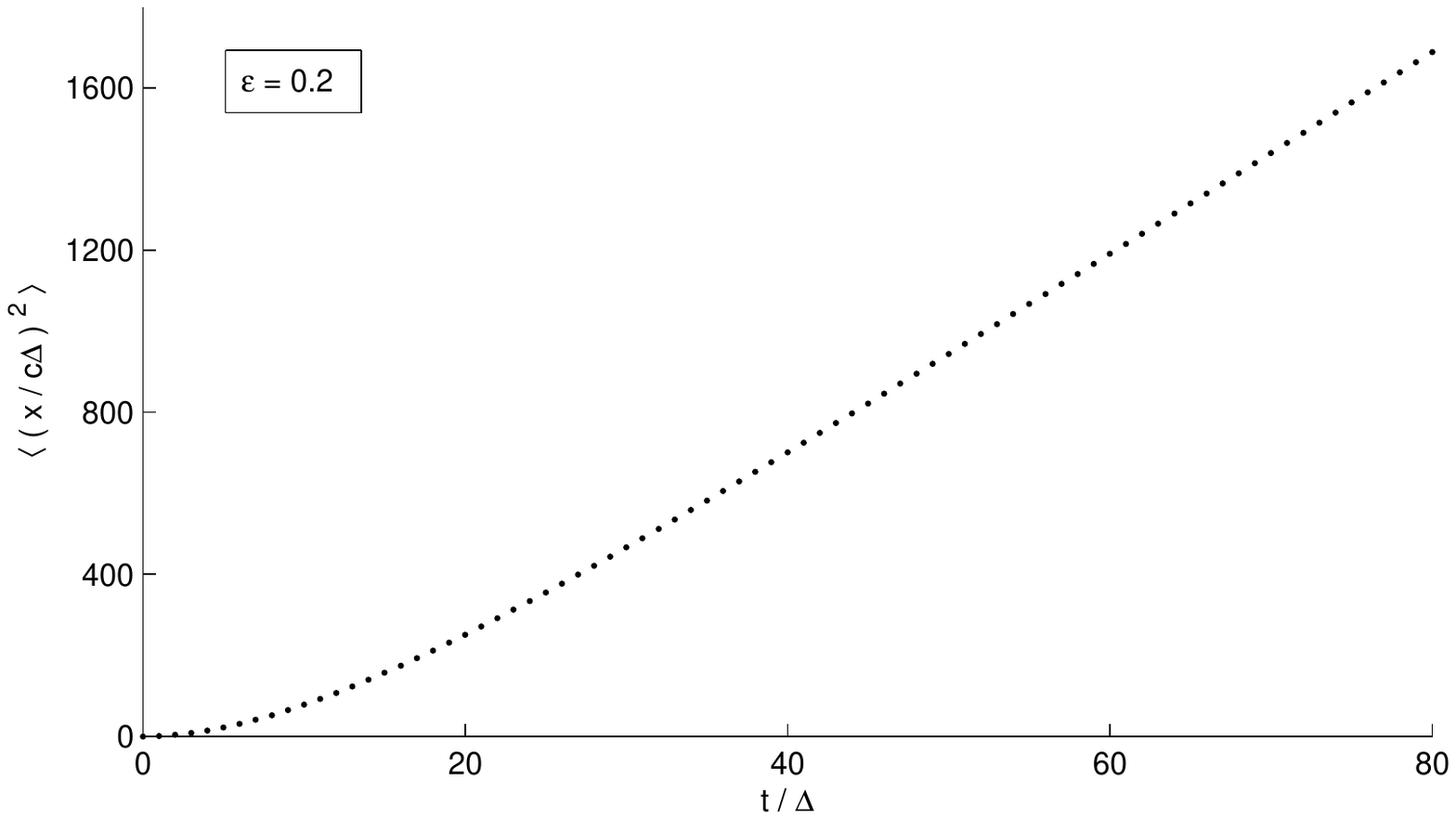}
\caption{Change from quadratic to linear growth of $\langle (x/c\Delta)^2\rangle$
with increasing 
$t=N\Delta$, in the case $\epsilon  =0.2$.  The transition occurs in the neighborhood of $N=25$, where $N\epsilon^2=1$. }
\end{figure}

The asymptotic behavior as $N\to\infty$ of the moments of the distribution \eqref{solutions2}, defined as
in \eqref{moments1},  has also been calculated \cite{BE3}, to give
\bea  
S_N^{(2n)}\sim \frac{N^n \,(2n-1)!}{(n-1)!\,2^{n-1}\epsilon^{2n}}\,,\qquad\qquad  
\mea\mea  
S_N^{(2n+1)}\sim \frac{ N^n\, (2n+1)!}{n!\,2^n\epsilon^{2n}}\frac{1-\epsilon^2}{2\epsilon^2}\,,  
\label{moments3}  
\eea  
for $n=0\,,1\,,2\,,\dots$. From this it follows that as $N\to\infty$, the probability distribution given by $P_{Nk}$ is asymptotic to a continuous, normal distribution with density 
\bea
P(N,k)=  \frac{e^{-\epsilon^2\,(k-k_{\,0})^2 /2N}}{\sqrt{2N\pi/\epsilon^2}}\,,\qquad -\infty<k<\infty\,, 
\label{asymptotic_density1}
\eea
where $k_0=(1-\epsilon^2)/2\epsilon^2$.  This generalizes the result \eqref{CRW1}, which is 
recovered when $\epsilon= 1$.  Fig. 4 shows the degree of agreement between the exact and asymptotic 
probability densities for $\epsilon=0.2$ with  $N=t/\Delta =100$ and  $N=300$; 
in the latter case, the plots sit almost one on top of the other.  Note that because the $P_{Nk}$ are defined
at $k$-values two units apart, we have for each value of $N$,  
\bea
\sum_{k=-\infty}^{\infty}\,\half P_{Nk}\,\Delta_k=1\,,\quad \Delta_k =2\,,
\label{dens_def}
\eea
showing that it is the set of values $\half P_{Nk}$, $k=-N\,,-N+2\,,\dots \,,N$ that is to be considered
a probability density for comparisons as in Fig. 4 with the continuous density $P(N,k)$.
\begin{figure}[t]
\centering
\includegraphics[width=10cm]{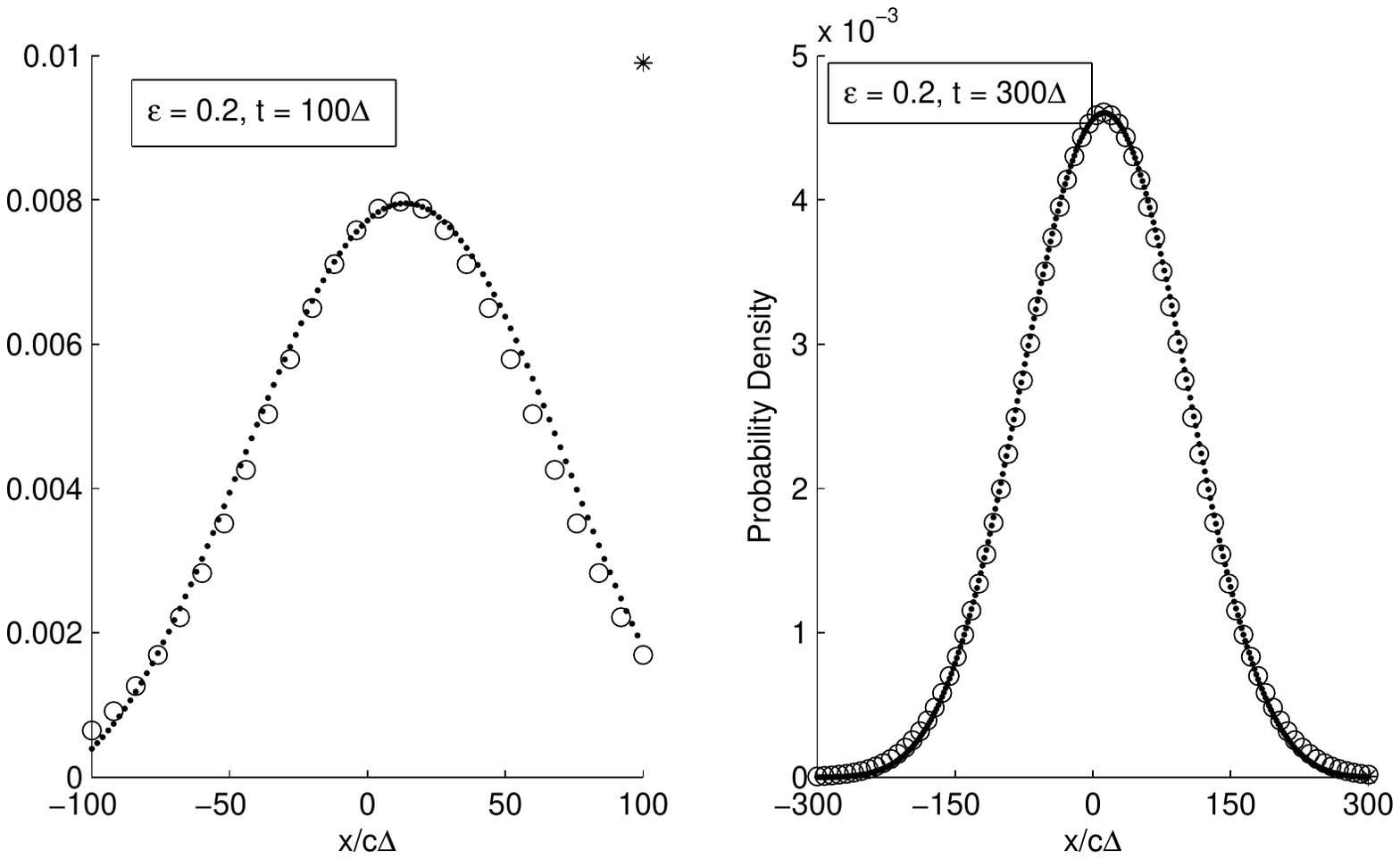}
\caption{Comparison of the discrete probability density $\half P_{Nk}$, $k=-N\,,-N+2\,,\dots\,,N$, marked
with dots, with the
asymptotic normal form $P(N,k)$, marked with circles, for $\epsilon=0.2$ and $N=100\,,300$.  
Note the value at $x=ct$ marked with an asterisk in the first plot. }
\end{figure}

For a general value of $\epsilon$, we now have that as $t=N\Delta\to \infty$,  
the probability density on the $x$-axis is asymptotic to that for the  normal distribution  
\bea  
{\cal P}(t,x)= \frac{e^{-(x-x_{\,0})^2/(4Dt)}}{\sqrt{4\pi Dt}}\,,\qquad -\infty<x<\infty\,,
\mea\mea
x_0=(1-\epsilon^2)c\Delta/2\epsilon^2\,,\quad D=c^2\Delta/2\epsilon^2  \,.  
\label{asymptotic_density2}  
\eea
Here the displacement of the mean position $\langle x\rangle$ of the particle at large times, 
to the value $x_0$, is noteworthy. 
Using \eqref{epsilon_def}, we see that this value is given in terms of the physical 
variables defining the Hamiltonian as
\bea
x_0=\frac{\cos(\mu B_0\Delta)}{1-\cos(\mu B_0\Delta)} \,c\Delta\,.
\label{displacement}
\eea
It can take any positive value, and is independent of the (very large value of the) time $t$.
We can also say that as $t\to\infty$,  there is a $95\%$ probability
of finding the particle within two standard deviations of the mean; this is the probability
to have $|x-x_0|<2c\sqrt{t\Delta} /\tan(\mu B_0\Delta/2)$.

\section{Discussion}
The transition from ballistic to diffusional behavior in QWs with unitary noise is surprising; in the
context of a relativistic quantum system as we have discussed here, it is even more remarkable. Diffusion
commonly arises as a non-relativistic, classical process.  Note from \eqref{epsilon_def} and 
\eqref{asymptotic_density2}
the dependence of the associated diffusion coefficient $D$ on the speed of light as well as 
on the magnetic moment, the strength of the magnetic field, and the time interval between 
changes in field direction.  

Can a system of the type we have described 
be realized physically?  There are several difficulties standing in the way.  
In the first place, it is now thought that all neutrinos have small rest masses, leaving us with no 
candidate massless Dirac particles.  Even if it should transpire that one of the neutrino types is 
after all massless, there is no evidence for neutrino magnetic moments, 
although there has long been speculation that such might exist \cite{cisneros,voloshin,pulido,barut,wong}.

Could the model apply to neutrinos, or indeed to the neutron, in situations where 
the rest-mass contribution $mc^2\beta$ now properly 
appearing as an extra term  in the  Hamiltonian \eqref{dirac1}, can nevertheless be neglected relative to the magnetic interaction term, so that \eqref{dirac1} still applies?   Evidently, this would require
\bea
|\hbar \mu B_0/2|\gg mc^2
\label{mass_term1}
\eea
in addition to the conditions \eqref{condition1}.  Inserting the observed values of $\mu$ and $m$ for the neutron
gives values of $|B_0|$ several orders of magnitude greater than any that have been 
achieved in the laboratory, although it is conceivable that such field strengths could occur in extreme 
cosmological situations.  Note that according to \eqref{condition1}, 
extremely large $|B_0|$ values require extremely  small 
field-oscillation times
$\Delta$  if our approximate treatment is to be valid.  In contrast to the situation for the free 
Dirac equation \cite{BES}, where the size of $\Delta$ can be adjusted at will 
to ensure accuracy of the approximations used there, in the present case it is determined 
once the external magnetic field is prescribed.   
  
For a neutrino (with anomalous magnetic moment) and very small rest mass, 
\eqref{mass_term1} is more easily satisfied. Although still unlikely in the laboratory, 
a process of the type we have described  
might apply in supernovas \cite{lychko} or in the solar corona, for example, and affect significantly  
the fluxes of one or more neutrino types through 
the very strong stochastic magnetic fields occurring there \cite{cisneros,voloshin,pulido,balantekin}.    

A more subtle difficulty that has been raised following \eqref{initial4} concerns the form of 
positive-energy states of
the relativistic particle.   Just as for a free electron \cite{flohr}, it is possible to 
construct positive-energy states of a free neutrino that 
are arbitrarily highly 
localized about any given point \cite{wood}, and we could have  used such a state 
in place of the $|\psi(0)\rangle$  in \eqref{start_vec1}, supposing that the particle 
is free at $t=0_-$, and that the 
interaction is switched on at $t=0_+$.  However, such a state and its translates do not have compact support 
and are not mutually orthogonal, severely complicating the analysis of the QW.  
Such a more complicated analysis is not warranted, in our opinion; for even  
if the particle is in a positive-energy state at $t=0_-$, it will not be so 
at $t=0_+$ when the field is switched on, because the Hamiltonian, and its positive energy states, 
change form.  
Similarly, even if it can be arranged that the particle is in a positive-energy state at 
$t=N\Delta_-$, it will not be so at $t=N\Delta_+$  
if the field abruptly changes direction at that time, for the same reason. 
This difficulty, which is perhaps related to the 
Klein paradox \cite{klein},  is not peculiar to the system we have discussed here, nor to our way of 
treating it.  It seems clear that it must beset the analysis of 
a massless particle in any time-dependent, 
discontinuously changing external field. One could try to overcome the difficulty by projecting 
onto positive energy states immediately after each discontinuity occurs in the Hamiltonian, 
but it is not clear how to do this in a simple way that preserves the unitarity of the 
resulting time evolution and hence the length
of the state vector.  

We have not attempted to resolve this difficulty here, contenting ourselves with analyzing 
the model as described, in the  belief that it makes an interesting addition to the 
collection of relativistic quantum systems that have been considered previously, and suggests that
a quite new type of behavior can appear in such systems when subject to stochastically 
varying external fields.  
We hope that it provokes further study.  

It would be interesting, and more realistic, to consider the Hamiltonian \eqref{dirac1} for 
the 
neutron, with rest-mass term added, and evaluate the evolution numerically, without making any
approximations like \eqref{evolution2}, to see if the stochastic nature of the interaction 
term continues to lead to
diffusional behavior at large times. However, the second difficulty mentioned above 
would still require resolution.  

The exact evolution in the massless case, associated with \eqref{evolution1}, could also be treated numerically to indicate any limitations of our approximate treatment. 

Finally, it should be mentioned that the present model can also be considered
in the context of quantum simulations of  relativistic effects
using the experimental apparatus of trapped ions. Indeed simulations of
Dirac's equation and associated  relativistic quantum effects for a single
trapped ion have already been proposed \cite{Zth} and experimentally 
realized \cite{Zexp}, especially concerning the simulation of the  Zitterbewegung phenomenon
and the Klein paradox  \cite{Kth,Kexp}. Simulations  of a Dirac particle in a magnetic potential 
and its topological
properties, using trapped ions,  have also been proposed  \cite{Dmagnf}. 
It is conceivable that the machinery of trapped ions might also allow a simulation of the
Dirac QW driven by fluctuating magnetic fields that we have developed above, although it would be
challenging for such simulations to overcome the  physical
obstacles indicated. 
\vs\ni
{\em Acknowledgment:}  We thank the School of Mathematics and Physics, 
University of Queensland, for its hospitality
during visits by D.E. (sabbatical) and I.S., when this work was completed.

\end{document}